\documentclass[
aps,%
12pt,%
final,%
notitlepage,%
oneside,%
onecolumn,%
nobibnotes,%
nofootinbib,%
superscriptaddress,%
noshowpacs,%
centertags]%
{revtex4}
\usepackage{amssymb}
\usepackage{epsfig}
\usepackage{multirow}
\usepackage{caption}
\usepackage{subfigure}
\usepackage{graphicx}
\usepackage{color}
\usepackage{amsmath}
\usepackage{cases}
\usepackage{amsthm}

\usepackage{array}  
\usepackage{makecell}

\usepackage{algorithm}
\usepackage{algorithmic}
\usepackage{booktabs}
\usepackage{url}
\usepackage{array, makecell, booktabs}

\theoremstyle{definition}

\theoremstyle{remark}

\def\bq{\begin{equation}}
\def\eq{\end{equation}}
\def\br{\begin{eqnarray}}
\def\er{\end{eqnarray}}
\def\brr{\bq\begin{array}{rlll}}
\def\err{\end{array}\eq}

\makeatletter
\def\serieslogo@{}
\makeatother \makeatletter
\def\@setcopyright{}
\makeatother

\begin{document}

\title{Multi-layer 5D Optical Data Storage: Mathematical Modeling and Deep Learning-Based Reconstruction of Birefringent Parameters}

\author{\firstname{Ye}~\surname{Zhang}}
\email{ye.zhang@smbu.edu.cn}
\affiliation{School of Mathematics and Statistics, Beijing Institute of Technology, 100081 Beijing, China}%

\author{\firstname{Qiao}~\surname{Zhu}}
\email{qzhu@bit.edu.cn}
\affiliation{School of Mathematics and Statistics, Beijing Institute of Technology, 100081 Beijing, China
}%

\author{\firstname{Rongkuan}~\surname{Zhou}}
\email{suzeat@foxmail.com}
\affiliation{MSU-BIT-SMBU Joint Research Center of Applied Mathematics, Shenzhen MSU-BIT University, 518172 Shenzhen, China}

\author{\firstname{Tatiana}~\surname{Lysak}}
\email{lysak@cs.msu.su}
\affiliation{Lomonosov Moscow State University, 119992 Moscow, Russia
}%

\author{\firstname{Chao}~\surname{Wang}}
\email{wangchao@smbu.edu.cn (corresponding_author)}
\affiliation{MSU-BIT-SMBU Joint Research Center of Applied Mathematics, Shenzhen MSU-BIT University, 518172 Shenzhen, China
}%


\begin{abstract}
Five-dimensional (5D) optical data storage has emerged as a promising technology for ultra–high-density, long-term data archiving. However, its practical realization is hindered by noise and interference during data readout. In this work, we develop a high-precision mathematical model for multi-layer 5D optical storage, grounded in the Jones matrix framework, to accurately capture polarization transformations induced by stacked birefringent nanostructures. Building on this model, we propose a 20-frame FiLM-conditioned U-Net algorithm to reconstruct birefringence parameters~--specifically, slow-axis orientation and retardance magnitude—directly from measured intensity patterns. Trained on both ideal and noisy datasets, the network demonstrates robust reconstruction performance under challenging measurement conditions. Compared with conventional frame-based retrieval approaches, our method achieves over an order-of-magnitude improvement in reconstruction accuracy. The proposed model and algorithm can be readily integrated into existing 5D optical readout systems, offering both a solid theoretical foundation and practical tools for precise data recovery.
\end{abstract}

\maketitle

\section{Introduction}

Five-dimensional (5D) optical data storage has recently emerged as a promising solution for ultra-high-density and long-term data archiving~\cite{zhang2014seemingly, malinauskas2016ultrafast, hu2021reversible, lu2022research, ren2023dual}. It encodes information using three spatial coordinates $(x,y,z)$, combined with two optical parameters derived from laser-induced nanostructures~\cite{sudrie1999writing, shimotsuma2003self, zhang2016eternal, shribak2003techniques}. Specifically, femtosecond laser pulses generate birefringent nanogratings inside fused silica. Thus information is encoded on each data voxel through its position, the slow-axis orientation, and the magnitude of optical retardance. This five-dimensional encodeing scheme enables extraordinary storage capacity and durability. For example, a single nanostructured quartz disc can theoretically store the order of $10^2\sim 10^3$ terabytes while maintaining stability at temperatures up to $1000^\circ C$~\cite{zhang2014seemingly}, making it highly suitable for long-term archival storage.

However, practical realization of 5D optical storage remains challenged by noise and interference during data readout~\cite{kazansky2020virtually, fedotov2021laser, r11, sakakura2020ultralow}. This significantly affects the accurate reconstruction of the encoded optical parameters. One major issue is the interlayer crosstalk. As voxel layers are packed closer together to increase storage density, signals from adjacent layers overlap, reducing decoding accuracy~\cite{ma2025error}. 
Furthermore, signal attenuation occurs as the probe beam travels through multiple layers, with cumulative absorption and scattering diminishing signal quality. Besides, additional degradation also arises from external noise sources, including detector fluctuations, background light, and instability in laser polarization. As a result, these factors lead to a complex, ill-posed inverse problem, where small measurement errors can result in large inaccuracies in retrieving voxel-specific optical parameters. This often forces compromise in encoding strategies, such as limiting the range of birefringent values to ensure decoding reliability. Altogether, robust and accurate parameter recovery from noisy and interfered measurements remains a central challenge for scalable 5D optical memory systems.

The core technical problem of reading 5D optical storage can be formulated as the inversion of polarization-resolved intensity data. Its goal is to reconstruct the slow-axis direction $\psi$ and retardance magnitudes $\Delta$ for each voxel. 
 This inversion is highly sensitive to noise and interlayer interference, especially in high-density configurations. 
 Therefore, accurate information retrieval requires an inverse reconstruction method that incorporates the underlying physics of light propagation, as well as a mechanism to suppress or compensate for noise.

Deep learning \cite{adler2017solving, seo2019learning, jin2017deep, li2020nett} offers a compelling approach to this challenge. Unlike traditional reconstruction methods that depend on explicit physical models, deep neural networks can learn complex mappings directly from data. This data-driven paradigm has proven effective in various ill-posed inverse problems, including image reconstruction and computational optics~\cite{barbastathis2019use, zhang2020review, ben2021deep, sinha2017lensless, wang2018image}. In optical storage applications, recent works have shown encouraging results. Shimobaba et al. \cite{shimobaba2017convolutional} utilized convolutional neural networks to read holographic memory. Similarly, Wiecha et al. \cite{wiecha2019pushing} applied machine learning to decode high-density planar nanostructures with minimal probing. Yang et al. \cite{yang2023high} demonstrated that deep networks can accurately reconstruct data from 3D hybrid nanostructures under noisy spectral conditions. In addition to these contributions, the potential of deep learning has recently been demonstrated in the classification and inverse design of nanoparticles \cite{jo2017holographic, malkiel2018plasmonic, peurifoy2018nanophotonic}. 
While earlier research mainly focused on single-layer storage~\cite{kuian2018accuracy, shribak2003techniques}, achieving the full capacity of 5D optical memory requires reliable readout across multiple closely spaced layers. Stacking layers greatly increases data density, but also leads to stronger interference between layers. This makes it necessary to develop accurate physical models and more reliable decoding methods. Therefore, building a robust and noise-tolerant reconstruction framework is the key to practical use of multi-layer optical storage.

In this work, we address the core inverse problem in 5D optical storage by developing a physics-informed deep learning framework. First, we derive a multilayer optical forward model based on the Jones matrix formalism, which describes how polarized light propagates through stacked birefringent nanogratings and how interlayer interference emerges. This model serves as the foundation for generating training data and understanding signal distortions. Second, we design a lightweight deep neural network. The network is based on the U-Net architecture~\cite{ronneberger2015u} and integrates Feature-wise Linear Modulation (FiLM) layers~\cite{perez2018film}. FiLM is a general conditioning mechanism that allows us to use physical prior information—specifically, the depth index $k$ of the layer being characterized—to dynamically modulate the behavior of network during feature extraction. This design enables our model to adaptively process signals originating from different depths, allowing for the direct reconstruction of the $\psi$ and $\Delta$ fields from noisy, overlapping intensity measurements. Our model is trained on simulated data incorporating realistic noise and interference conditions, enabling it to learn robust feature mappings. Through extensive numerical experiments, we demonstrate that the proposed method achieves high accuracy and robustness in retrieving birefringence parameters, even under dense and noisy conditions.

The remainder of this paper is organized as follows. Section~\ref{sec:models} introduces the physical model of optical propagation and defines the interference metric. Section~\ref{sec:InverseAlg} describes the inverse reconstruction algorithm and training procedure. Section~\ref{Sec:Num} presents numerical results validating the approach. Finally, Section~\ref{Sec:Con} concludes with a discussion of future research directions in 5D optical data storage. 

\section{5D Optical Storage Principles and Mathematical Modeling}\label{sec:models}

5D optical storage based on fused quartz employs femtosecond laser pulses to induce nanogratings within the medium. These nanogratings induce form birefringence, enabling each voxel to encode information in five independent degrees of freedom. Specifically, two optical parameters, i.e., the slow-axis orientation and the magnitude of optical retardance, can be independent controlled. The slow-axis orientation is governed by the polarization state of the writing beam, while the retardance is modulated via adjusting the laser pulse energy. This independent tunability enables each voxel to represent up to 16 distinct states, thereby encoding 4 bits of information per voxel unit.
\begin{figure}[htbp] \centering 
\includegraphics[width=0.9\linewidth]{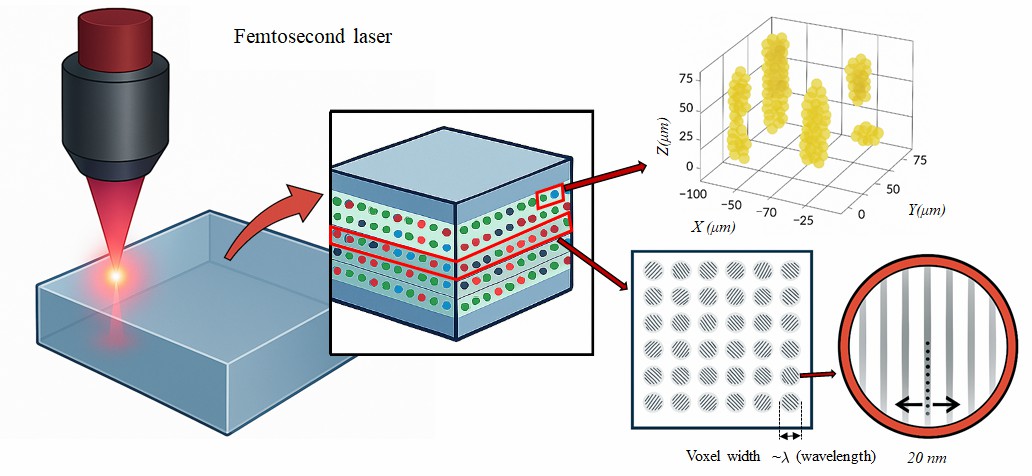} 
\caption{Schematic illustration of 5D optical data storage based on femtosecond laser nanostructuring in fused silica.} \label{fig:ML5D} 
\end{figure}

FIG.~\ref{fig:ML5D} outlines the entire process of 5D optical data writing. A focused femtosecond laser beam generates spatially localized birefringent nanostructures within the quartz substrate. Thus data is stored in multilayer voxel arrays arranged in depth, with each voxel containing periodic nanogratings. These nanogratings, with sub-micron periodicity on the order of~$20\mu m$, provide the basis for ultra-dense data encoding. The 3D spatial distribution of voxels illustrates the precise stacking capability that enables multilayer storage with micrometer-level interlayer spacing.

\begin{figure}[htbp]
\centering
\includegraphics[width=1.0\linewidth]{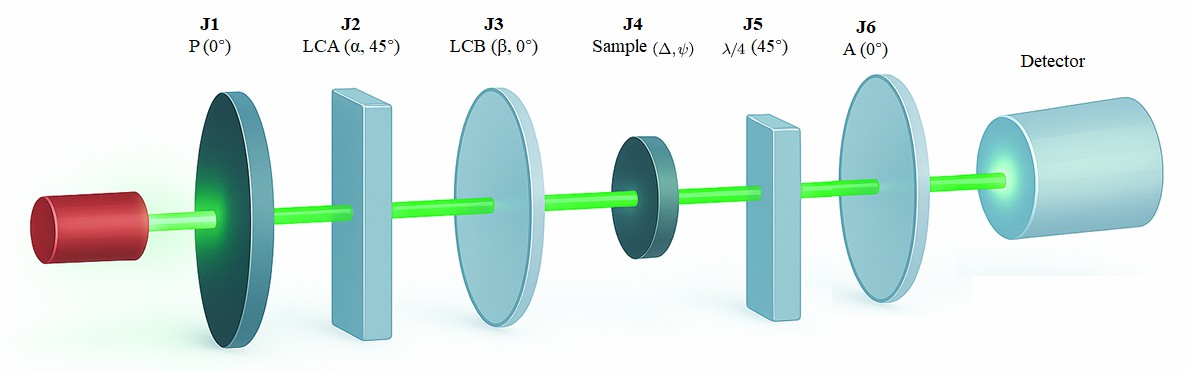}
\caption{Schematic diagram of the 5D optical writing and detection system.}
\label{fig:optics}
\end{figure}

The readout process for 5D storage relies on the birefringence effects induced by nanogratings. 
To extract the stored information, it is necessary to measure the polarization response of the transmitted light and extract the two key birefringence parameters: the slow-axis direction $\psi(x,y)$ and the retardance magnitude $\Delta(x,y)$. The optical detection system used for readout is illustrated in FIG.~\ref{fig:optics}. It inclueds a coherent light source, a linear polarizer (P), two liquid crystal variable retarders (LCA and LCB). The light transmitted through the sample is analyzed using a circular polarization analyzer composed of a quarter-wave plate and a second linear polarizer, before being captured by a CCD or CMOS sensor. By varying the control parameters of LCA and LCB and recording the corresponding transmitted intensity, the system reconstructs the birefringence maps $\left(\psi(x,y), \Delta(x,y)\right)$ for each imaging plane.

In multilayer storage configuration, the same detection system is employed, but the focal plane is adjusted to target different depth layers within the medium. 
However, due to the limited axial resolution of the optical system, the signal at a given focal depth inevitably contains contributions from out-of-focus voxels in neighboring layers, leading to interlayer crosstalk. Additional aberrations may also arise from imperfections in optical alignment. Consequently, the acquired signal represents a composite of multiple voxel layers, necessitating robust inversion algorithms to decouple overlapping contributions and accurately recover voxel-level birefringence.

\subsection{Mathematical model}
To characterize the optical readout process of 5D birefringent storage, we employ Cartesian coordinates $(x,y)$ to describe the distribution of the retardance magnitudes $\Delta(x,y)$ and the slow-axis orientation $\psi(x,y)$ within the image plane. The polarization-sensitive elements in the optical system, including polarizers, liquid crystal variable retarders (LCVRs), the birefringent sample, and a quarter-wave plate, are modeled using Jones calculus. By sequentially multiplying the corresponding Jones matrices of these components, the resulting transmitted light intensity $I(\alpha, \beta, x, y)$ at each point on the detector is analytically expressed as follows~\cite{shribak2003techniques}:
\begin{equation}\label{eq:computeIwithoutnoise}
    \begin{aligned}
        I(\alpha,\:&\beta,\:x,\:y)= \frac12\tau(x,y)I_{\max}(x,y)[1+\sin\alpha\cos\beta\cos\Delta(x,y) \\
        &-\sin\alpha\sin\beta\cos2\psi(x,y)\sin\Delta(x,y) +\:\cos\alpha\:\sin2\psi(x,y)\sin\Delta(x,y)] +I_{\min}(x,y).
    \end{aligned}
\end{equation}
where $\alpha$ and $\beta$ denote the phase modulation angles introduced by LCA and LCB, respectively. The terms $I_{max}(x,y)$ and $I_{min}(x,y)$ correspond to the modulated signal intensity and background illumination, respectively. The factor $\tau(x,y)$ denotes the distribution of the isotropic transmittance of the medium. The retardance value $\Delta(x, y)$ ranges from $0$ to $\pi$. Furthermore, we constrain the slow-axis orientation $\psi(x,y)$ to the range $[0,\pi/2]$. Although birefringent materials allow $\psi$ to vary in the interval $[0,\pi]$~\cite{shribak2003techniques}, values in the range $(\pi/2,\pi]$ are indistinguishable from those in $[0,\pi/2]$ under traditional intensity-only Jones-based inversion. 
To ensure consistency and facilitate quantitative comparisons across reconstruction algorithms, we limit $\psi$ to $[0,\pi/2]$ in all subsequent experiments and simulations.
For simplification, we assume $\tau(x,y)$ to be a constant scalar $\tau$, which is justified in our setup due to the spatial homogeneity of the transparency for sample, i.e., angular scattering effects and out-of-plane anisotropies are neglected. 


This forward model serves as the theoretical foundation for the subsequent inverse problem, where the birefringent parameters $\Delta$ and $\psi$ are reconstructed from intensity measurements under various polarization modulation settings $(\alpha, \beta)$.

\subsection{Multi-layer model with noise}
In multilayer 5D optical storage systems, the readout beam propagates through multiple recording layers before reaching the target voxel, introducing various forms of signal distortion through various physical processes. These effects are collectively modeled as noise components, classified into three principal types.
\begin{itemize}
    \item[1.] \textbf{Parameter jitter noise}:\\
    During laser writing, the desired birefringence parameters may deviate from their desired values due to intrinsic fluctuations such as thermal drift, spectral jitter, or polarization instability. These uncertainties are modeled as additive Gaussian noise affecting the retardance $\Delta$ and slow-axis angle $\psi$:
    \begin{align*}
        \Delta=\Delta^{\text{true}}+\delta_\Delta,\quad \psi=\psi^{\text{true}}+\delta_\psi,\quad \text{with}\quad \delta_\Delta\sim\mathcal{N}(0,\sigma^2_{\Delta}),\quad \delta_\psi\sim\mathcal{N}(0,\sigma^2_{\psi}).
    \end{align*}
    Such variations result in mismatches between the intended and actual birefringent states, which propagate into the intensity measurements and degrade decoding accuracy. 
    \item[2.] \textbf{Multiplicative intensity noise}:\\
    As the light beam traverses successive birefringent layers, its intensity can be modulated by inter-layer interactions and fluctuations in the light source. Firstly, temporal instability of the light source output leads to variations in the detected peak intensity. This effects is modeled as follows:
    \begin{align*}
        I_{\text{max}}^{\text{obs}}=I_{\text{max}} \cdot (1+\varepsilon),\qquad \text{with}\qquad\varepsilon\sim\mathcal{N}(0,\sigma^2_{\text{src}}). 
    \end{align*}
    Secondly, coherent interactions between nonogratings in different layers produce additional multiplicative fluctuations. They represent one of the dominant forms of noise and can be partially mitigated via optical path separation. A rigorous derivation for the inter-layer effect will be given in the rest of this section. 
    \item[3.] \textbf{Additive background noise}:\\
    In addition to the above factors, readout measurements are affected by extrinsic additive noise, arising from dark current in the detector, ambient stray light, and optical reflections. This component is modeled as:
    \begin{align*}
        I_{\text{add}} \sim \mathcal{N}(0,\sigma^2_{\text{add}}).
    \end{align*}
    Additive noise is independent of the stored birefringence values but accumulates as system complexity increases, especially under suboptimal environmental or hardware conditions. Besides, background light $I_{\text{min}}$ also contributes a constant offset fluctuation to the measured signal.
\end{itemize}

To rigorously describe the optical response of the multilayer birefringent medium, we begin by formulating the light propagation process using Jones calculus. Let $\boldsymbol{E}_0(x,y)$ denote the incident electric field, expressed as a monochromatic plane wave:
\begin{align}
    \boldsymbol{E}_0(z,t)=\begin{bmatrix}
        E_x \\
        E_y
    \end{bmatrix}\text{exp}\left(\frac{2\pi i}{\lambda}(z-ct)\right),
\end{align}
where $\lambda$ is the wavelength of the light, and $E_x,\, E_y$ represent the amplitude components along the $x$ and $y$ axes.
In the current system, a linearly polarized coherent beam is employed, with horizontal polarization aligned along the $x$-axis, such that: $\begin{bmatrix}
        E_x \quad E_y
    \end{bmatrix}^T=\begin{bmatrix}
        1 \quad 0
    \end{bmatrix}^T$. 

The beam subsequently passes through a series of optical components. These components include linear polarizers and birefringent phase retarders, each represented by a corresponding Jones matrix. As illustrated in FIG.~\ref{fig:optics}, the beam sequentially interacts with six principal optical elements in the following order: (1) a linear polarizer $\boldsymbol{J}_1=\begin{bmatrix}
    1 & 0 \\
    0 & 1
\end{bmatrix}$ aligned along the horizontal axis; (2) a liquid crystal variable retarder (LCA) $\boldsymbol{J}_2=\boldsymbol{J}_r(\alpha,\pi/4)$; (3) a second liquid crystal variable retarder (LCB) $\boldsymbol{J}_3=\boldsymbol{J}_r(\beta,0)$; (4) a birefringent specimen $\boldsymbol{J}_4=\boldsymbol{J}_r(\Delta(x,y),\psi(x,y))$; (5) a quarter-wave plate $\boldsymbol{J}_5=\boldsymbol{J}_r(\pi/4,\pi/4)$; and (6) a second linear polarizer $\boldsymbol{J}_6=\boldsymbol{J}_1$ serving as the linear analyzer. Each $\boldsymbol{J}_r(\psi,\theta)$ denotes a Jone matrix of a phase retarder with retardance $\psi$ and optical axis oriented at angle $\theta$, constructed as follows:
\begin{align*}
    \boldsymbol{J}_r(\psi,\theta)=R(-\theta)\cdot \begin{bmatrix}
        e^{-i\psi/2} & 0 \\
        0 & e^{i\psi/2}
    \end{bmatrix}\cdot R(\theta), \qquad R(\theta)=\begin{bmatrix}
        \cos\theta & -\sin\theta \\
        \sin\theta & \cos\theta
    \end{bmatrix}.
\end{align*}
The overall transformation of the incident field is governed by the cumulative Jones matrix, and the output field $\boldsymbol{E}_{\text{out}}$ is given by:
\begin{align}\label{eq:Eout}
    \boldsymbol{E}_{\text{out}} = \boldsymbol{J}_6\cdot\boldsymbol{J}_5\cdot\boldsymbol{J}_4\cdot\boldsymbol{J}_3\cdot\boldsymbol{J}_2\cdot\boldsymbol{J}_1\cdot \boldsymbol{E}_0.
\end{align}
The output intensity detected at the image plane is proportional to the squared magnitude of the resulting Jones vector. Specfically:
\begin{align}
    I_{\text{out}}=C|\!|\boldsymbol{E}_{\text{out}}|\!|^2=C\boldsymbol{E}^*_{\text{out}}\cdot\boldsymbol{E}_{\text{out}},
\end{align}
where $C$ is a proportionality constant that depends on the detector response function, the optical path efficiency, and any system-specific gain factors. In numerical simulations or normalized models, we set $C=1$ for simplicity.

In multilayer configurations, the probe beam encounters multiple birefringent layers prior to reaching the layer of interest. These preceding layers each induce a polarization transformation and collectively modify the light field. To capture this cumulative effect, the sample matrix $\boldsymbol{J}_4$ in Eq.~\eqref{eq:Eout} is extended to:
\begin{align}
    \boldsymbol{J}_4=\hat{\boldsymbol{J}}^N_4\cdot\hat{\boldsymbol{J}}^{N-1}_4\cdots\boldsymbol{J}_r\left(\Delta+\delta_\Delta,\psi+\delta_\psi\right)\cdot\hat{\boldsymbol{J}}^{k-1}_4\cdots\hat{\boldsymbol{J}}^1_4,
\end{align}
where each $\hat{\boldsymbol{J}}^i_4\,\, (1\leq i\leq N)$ denotes the Jones matrix of the $i$-th birefringent layer along the light path, and parameters $\delta_\Delta$, $\delta_\psi$ model stochastic perturbations (e.g., jitter, fabrication error) in retardance and axis orientation. This formulation accounts for the accumulated modulation caused by all intervening voxels and serves as the basis for forward modeling under noisy, multilayer conditions.

\begin{figure}[htbp]
\centering
\includegraphics[width=0.85\linewidth]{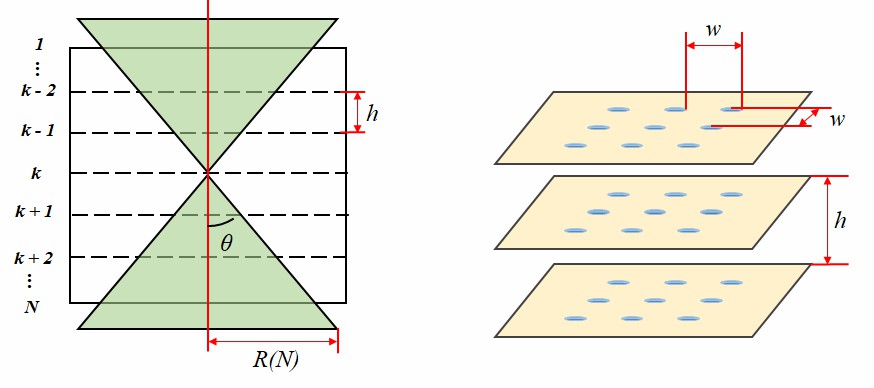}
\caption{Multi-layer 5D optical storage model.}
\label{fig:MulLayer}
\end{figure}

To quantitatively describe voxel-level crosstalk in multilayer 5D optical storage, a set of simplifying assumptions is introduced to make the modeling tractable while retaining physical fidelity. First, each voxel is modeled as a planar birefringent element with uniform properties in the axial direction. That is, vertical (height) variations of the nanogratings within a single layer are neglected, while the axial spacing between layers is preserved. Second, due to the geometric complexity and variability in the spatial arrangement of nanogratings across the sample volume, the influence of out-of-focus layers is approximated statistically. Specifically, the birefringence modulation from non-target layers is assumed to be spatially uniform and random, thereby avoiding explicit modeling of voxel-to-voxel interactions. Third, it is assumed that during readout, the probe beam sequentially traverses a total of $N$ storage layers, with the focal plane positioned on the target layer. Each layer encountered along the optical path imposes birefringence-induced modulation on the beam, which accumulates and affects the final intensity recorded at the detector. As shown in FIG.~\ref{fig:MulLayer}, a collimated probe beam is focused onto the target layer, passing through all preceding and subsequent layers.

Due to the focusing of the readout beam onto the target layer, the cross-sectional area of the light path varies with depth. Layers positioned farther from the focal plane intersect a wider beam footprint, thereby encompassing a larger number of nanogratings. Consequently, these out-of-focus layers exert a stronger cumulative modulation on the beam despite their distance from the focal plane. In this model, the effective contribution of each layer is thus governed by both its birefringence properties and its geometric distance from the focal plane.
%

To quantitatively evaluate the number of nanogratings intersected by the probe beam in a given storage layer, we derive an analytical approximation based on geometric considerations. Each storage layer is assumed to contain nanogratings arranged in a regular square lattice with constant lateral pitch $\omega$ and uniform interlayer spacing $h$, as illustrated in FIG.~\ref{fig:MulLayer}. The optical readout beam is modeled as a circular cone with semi-aperture angle $\theta$, and is focused on a specific target layer indexed by $k$ ($1\leq k\leq N$). For any layer indexed by $n$ ($1\leq n\leq N$), the radius of the beam footprint at that layer, i.e., intersection area of the lateral extent of the probe beam, is given by:
\begin{align}
    R(n)=|n-k|\cdot h\cdot\tan\theta, \quad n=1, \cdots, N.
\end{align}
This radius determines the lateral extent of the intersection area within the $n$-th layer. 

Let $r = R(n)/\omega$ denote the normalized radius in units of the grating pitch. We estimate the number of discrete nanogratings intersected by the beam using a circular approximation on the underlying integer lattice. The total number of grid points (nanogratings) within this region can be decomposed into three contributions:
\begin{itemize}
    \item[1.] \textbf{Central point}: the beam center always intersects the nanograting located at the origin, contributing 1.
    \item[2.] \textbf{Points along the coordinate axes}: for each integer $i$ from 1 to $\lfloor r \rfloor$, four symmetric points lie along the axes, contributing $4\lfloor r \rfloor$.
    \item[3.] \textbf{Interior lattice points}: for each $i=1,\cdots,\lfloor r\rfloor$, the maximum integer $j$ such that the point $(i,j)$ lies inside the circle satisfies:
    \begin{align*}
        i^2+j^2\leq r^2\qquad \rightarrow \qquad j\leq \sqrt{r^2-i^2}.
    \end{align*}
    Hence, for each $i$, there are $\lfloor\sqrt{r^2-i^2}\rfloor$ valid $j$ values, and by symmetry this contributes $4$ times the sum over $i$ to cover all four quadrants.
\end{itemize}
Combining these terms, the total number of nanogratings, denoted as $L(n)$, intersected by the beam at the $n$-th layer is approximated by:
\begin{equation}
L(n) = 
\begin{cases}
1 + 4 \left\lfloor r \right\rfloor 
+ 4 \displaystyle\sum_{i=1}^{\left\lfloor r \right\rfloor} 
\left\lfloor \sqrt{r^2 - i^2} \right\rfloor, & n \neq k \\
1, & n = k
\end{cases}
\end{equation}
This analytical approximation provides a layer-wise estimate of the number of nanogratings modulating the light beam, which is crucial for modeling cumulative inter-layer interference in multilayer optical storage. The result can be used as a weighting factor for estimating noise contributions from each layer during birefringence-based signal recovery.

To further quantify the modulation effect of each intermediate storage layer, we introduce the grating density parameter $d$, which denotes the area fraction covered by birefringent structures within each layer. Each nanograting is modeled by a random Jones matrix $\boldsymbol{J}_g$, representing the stochastic birefringent induced by the local anisotropy. Furthermore, to account for the distance-dependent impact of each layer on the probe beam, we introduce a layer-specific weighting factor $\sigma(n)$, which decreases with increasing axial distance from the target layer. This reflects the diminishing influence of distant layer due to beam focusing and attenuation. Thus, the cumulative modulation of all nanogratings in layer $n$ can be modeled as:
\begin{equation}
\hat{\boldsymbol{J}}^n_4 = \boldsymbol{I} + \sigma(n)\cdot \frac{d}{L(n)} \sum_{g=1}^{L(n)} \boldsymbol{J}_g, \quad n=1, \cdots, N, 
\end{equation}
where $\boldsymbol{I}$ is the identity matrix, representing the undisturbed polarization state in the absence of modulation. The summation term models the averaged modulation due to randomly distributed nanogratings in the layer, scaled by the grating density. 
%

The overall intensity measured by the detector under this cumulative modulation can then be written as:
\begin{align}\label{eq:cumpImax}
    I_{\text{noise}}(\alpha,\beta,\Delta,\psi,k)=(1+\varepsilon)\cdot|\!|\mathbf{J}_{\text{ps}}\cdot \mathbf{E}_0|\!|^2+I_{\text{min}}+I_{\text{add}},
\end{align}
where the total system matrix $\mathbf{J}_{\text{ps}}$ is defined as:
\begin{equation}
\boldsymbol{J}_{\text{ps}} = \boldsymbol{J}_6 \cdot \boldsymbol{J}_5 \cdot \underbrace{(\hat{\boldsymbol{J}}^N_4 \cdots \hat{\boldsymbol{J}}^1_4)}_{\boldsymbol{J_4}} \cdot \boldsymbol{J}_3 \cdot \boldsymbol{J}_2 \cdot \boldsymbol{J}_1.
\end{equation}
This formulation reflects the full optical path from the initial polarization control to the analyzer, incorporating all birefringent layers.


\section{Inverse reconstruction algorithms}\label{sec:InverseAlg}

In practical multi-layer 5D optical storage systems, the reliable reconstruction of birefringence parameters is significantly challenged by signal degradation resulting from interlayer crosstalk, detector noise, and imperfections in the optical model. Traditional analytical reconstruction techniques, such as four-frame and five-frame algorithms, are theoretically sound but highly sensitive to measurement noise and optical misalignment. As demonstrated in TABLE~\ref{tab:errors_with_and_without_noise} in Section~\ref{Results}, these methods often yield inaccurate or unstable estimates of phase retardation $\Delta$ and the optical axis orientation $\psi$ under realistic noisy conditions. 

\subsection{Analytical algorithms revisited}

Based on polarization-resolved detection configurations mapped on the Poincaré sphere, five distinct intensity measurements $\Sigma_i\,\, (i=0,\cdots,4)$ are acquired for each voxel. Specifically, $\Sigma_0$ corresponds to right-handed circular polarization, while $\Sigma_i$ $(i = 1, \cdots, 4)$ correspond to elliptical polarization states generated by modulating the liquid crystal retarders with an angular swing parameter $\chi$. The modulation settings for each measurement are given by:
\begin{align*}
    &\Sigma_0: \alpha=90^\circ, \beta=180^\circ,\qquad\qquad \Sigma_1: \alpha=90^\circ-\chi, \beta=180^\circ,\\
    &\Sigma_2: \alpha=90^\circ+\chi,\beta=180^\circ, \qquad\, \Sigma_3: \alpha=90^\circ, \beta=180^\circ-\chi, \\
    &\Sigma_4: \alpha=90^\circ, \beta=180^\circ + \chi.
\end{align*}
The parameter $\chi$ controls the ellipticity of the polarization states and is typically set around $10^\circ$ to enhance the sensitivity to small retardance values.
The intensity measurements under five elliptical polarization states are expressed as follows:
\begin{align}
    \begin{cases}
        I_0&=\frac12\tau I_{\max}[1-\cos\Delta]+I_{\min}, \\
        I_1&=\frac12\:\tau I_{\max}[1-\cos\chi\cos\Delta +\sin\chi\sin2\psi \sin\Delta] +I_{\min}, \\
        I_{2}&=\frac{1}{2}\:\tau I_{\max}[1-\cos\chi\cos\Delta -\sin\chi\sin2\psi\sin\Delta]+I_{\mathrm{min}},\\
        I_{3}&=\frac12\:\tau I_{\max}[1-\cos\chi\cos\Delta -\sin\chi\cos2\psi \sin\Delta ] + I_{\min}, \\
        I_{4}&=\frac12\tau I_{\max}[1-\cos\chi\cos\Delta +\sin\chi\cos2\psi \sin\Delta ] +I_{\min}
    \end{cases}
\end{align}

Based on the above intensity measurements, two birefringence parameters, i.e., retardance $\Delta$ and slow-axis orientation $\psi$, can be analytically recovered via different frame-based algorithms~\cite{shribak2003techniques}.


\textbf{Four-frame algorithm $(\Sigma_0,\Sigma_1,\Sigma_2,\Sigma_3)$:}
\begin{align}\label{FourFrameAlg}
    \begin{cases}
        &A=\frac{I_{1}-I_{2}}{I_{1}+I_{2}-2I_{0}}\tan\frac{\chi}{2} (=\sin2\psi\cdot\tan\Delta), B=\frac{I_{1}-I_{2}-2I_{3}}{I_{1}+I_{2}-2I_{0}}\tan\frac{\chi}{2} (=\cos2\psi\cdot\tan\Delta), \\ &
        \Delta=
        \begin{cases}\quad\arctan\left[(A^2+B^2)^{\frac{1}{2}}\right],I_1+I_2-2I_0\geq0, \\\pi-\arctan\left[(A^2+B^2)^{\frac{1}{2}}\right],I_1+I_2-2I_0<0,
        \end{cases}
        \\&\psi=\frac{1}{2}\arctan\left(\frac{A}{B}\right)+k\pi,\quad k=0, 1.
    \end{cases}
\end{align}

\textbf{Five-frame algorithm $(\Sigma_0,\Sigma_1,\Sigma_2,\Sigma_3, \Sigma_4)$:}
\begin{align}\label{FiveFrameAlg}
    \begin{cases}
        &A=\frac{I_{1}-I_{2}}{I_{1}+I_{2}-2I_{0}}\tan\frac{\chi}{2} (=\sin2\psi\cdot\tan\Delta), B=\frac{I_{4}-I_{3}}{I_{3}+I_{4}-2I_{0}}\tan\frac{\chi}{2} (=\cos2\psi\cdot\tan\Delta),\\ & \Delta=
        \begin{cases}\quad\arctan\left[(A^2+B^2)^{\frac{1}{2}}\right],I_1+I_2-2I_0\geq0, 
        \\\pi-\arctan\left[(A^2+B^2)^{\frac{1}{2}}\right],I_1+I_2-2I_0<0,
        \end{cases} \\ &
        \psi=\frac{1}{2}\arctan\left(\frac{A}{B}\right) +k\pi,\quad k=0, 1.
    \end{cases}
\end{align}
These analytical formulas enable direct recovery of linear birefringence parameters from intensity measurements and are particularly effective in ideal, low-noise scenarios. However, in practical high-density 5D optical storage environments, where multiple layers introduce signal blending and noise—their robustness degrades substantially. This motivates the need for more advanced, data-driven inversion strategies, as discussed in the following subsection.

\subsection{An efficient neural network inversion algorithm for multilayer model with noise}

To overcome these limitations, we propose a robust, data-driven inverse reconstruction approach based on deep learning. Our method replace analytical inversion with a supervised learning framework that directly learns the nonlinear mapping from observed intensity images to the underlying birefringence distribution. Specifically, we construct a neural network that takes as input a set of 20 modulation angles $\{(\alpha_i, \beta_i)\}$ and their associated intensity distribution $\{I_i\}$, along with a conditional parameter $k$ specifying the layer to be characterized. The network then outputs spatially resolved estimates of $\Delta^{\text{pred}} \in [0, \pi]$ and $\psi^{\text{pred}} \in [0, \pi/2]$.

\begin{figure}[htbp]
\centering
\includegraphics[scale=0.65]{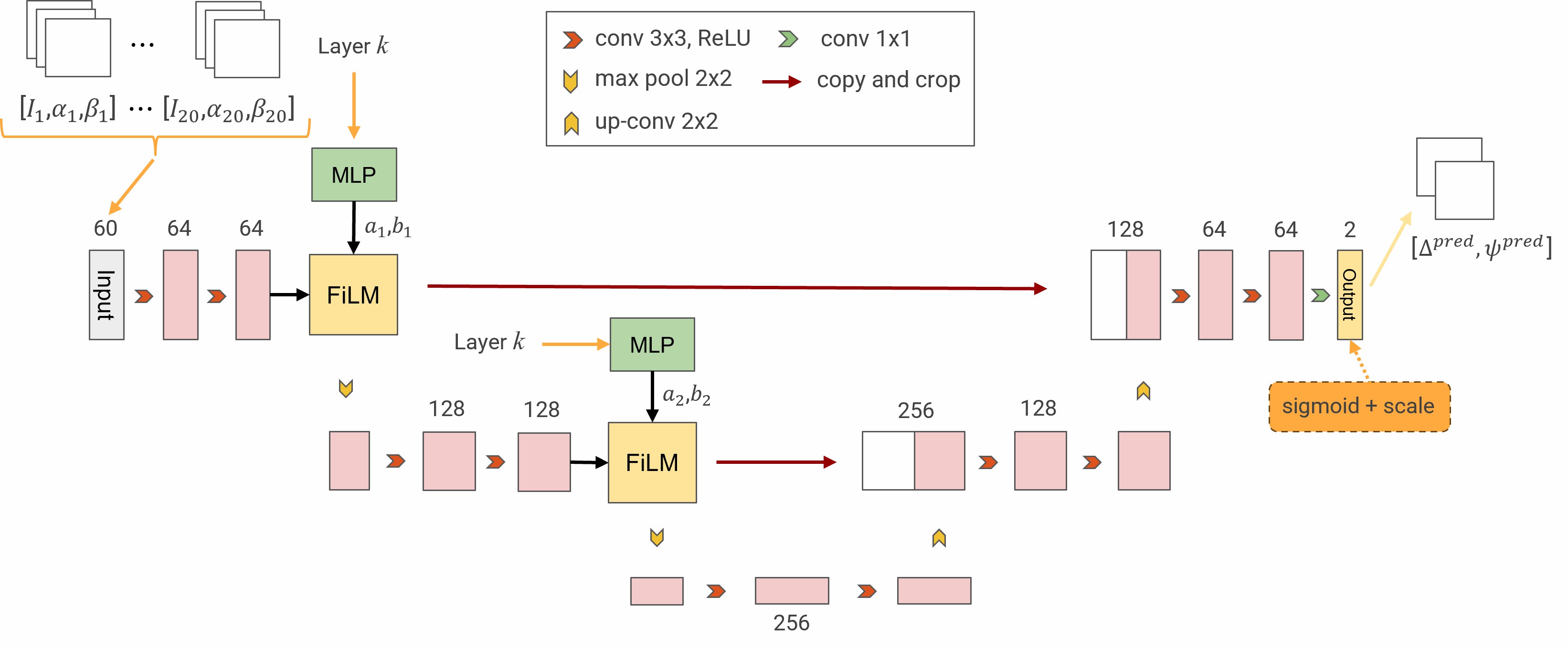}
\caption{20‑frame FiLM‑Conditioned U‑Net architecture}
\label{fig:network_architecture}
\end{figure}

The backbone of our approach is a U-Net architecture designed to process a 60-channel input tensor. This tensor is formed by concatenating 20 measurement sets, each represented as a 3-channel map comprising the intensity distribution $I_i$ and its corresponding modulation angles $\alpha_i$ and $\beta_i$, which are spatially broadcast to match the dimensions of $I_i$. This architecture is particularly well-suited to inverse problems involving spatially structured outputs. The network consists of an encoder-decoder configuration with symmetric skip connections that preserve fine-grained spatial features while capturing multiscale context. Crucially, our model incorporates physical context via a conditional mechanism based on FiLM. The layer index $k$ is first mapped to a dense vector via an embedding layer, which is then passed through the $k$-Conditioner—a small multi-layer perceptron (MLP) with two linear layers and ReLU activation. This module produces the FiLM scale ($a$) and bias ($b$) parameters. These parameters are then applied within FiLM layers at two stages of the encoder to adaptively modulate the feature maps. Specifically, at each modulation stage~$l$, the output feature map~$F_{l}$ from the previous convolution block is transformed as:
 $
\hat{F}_{l} = a_{l} \cdot F_{l} + b_{l}
$. This allows the network to dynamically adjust its feature extraction process based on the physical properties of the target layer. A final $1 \times 1$ convolution layer with two output channels produces a pair of spatial maps corresponding to the birefringence parameters. To ensure physical plausibility, a sigmoid activation is applied to constrain the outputs to $(0, 1)$. These normalized values are then linearly rescaled to the target intervals $[0, \pi]$ for $\Delta^{\text{pred}}$ and $[0, \pi/2]$ for $\psi^{\text{pred}}$, respectively. The architecture of the proposed U-Net model is illustrated in Fig.~\ref{fig:network_architecture}.

Unlike model-based approaches, our method does not require explicit knowledge of the forward imaging model or birefringence-induced phase relations. Instead, the network learns to approximate the inverse mapping in a data-driven manner, enabling accurate reconstruction even in scenarios with substantial noise or modeling uncertainty. The full procedure is summarized in Algorithm \ref{alg:supervised-inversion}, which outlines dataset synthesis, network configuration, and training strategy.

\begin{algorithm}[htbp]
\caption{20‑frame FiLM‑Conditioned U‑Net for estimating birefringenc parameters in multi-layer 5D optical data storage}
\label{alg:supervised-inversion}
\begin{algorithmic}[1]
\REQUIRE  
  Birefringence maps $\{(\Delta_j^{\text{true}},\psi_j^{\text{true}})\in \mathbb{R}^{H\times W}\}_{j=1}^{M}$,
  layer index $k\in\{1,\dots,L\}$,
  shared angles $\{(\alpha_i,\beta_i)\}_{i=1}^{20}$,
  training epochs $K$, validation ratio $\nu$, early‑stopping patience $P$
\ENSURE  
  Trained network $f_{\theta^*}:\bigl(\{(I_i,\alpha_i,\beta_i)\}_{i=1}^{20},k\bigr)\!\mapsto\!
                 (\Delta^{\text{pred}},\psi^{\text{pred}})$
\STATE\textbf{Data Generation}
\FOR{$j=1$ to $M$}
    \FOR{$i=1$ to $20$}
        \IF{$i\le10$}              
            \STATE $I_{j,i}\!\leftarrow\! I_{\text{ideal}}(\alpha_i,\beta_i,\Delta_j^{\text{true}},\psi_j^{\text{true}})$
        \ELSE
            \STATE $I_{j,i}\!\leftarrow\! I_{\text{noisy}}(\alpha_i,\beta_i,\Delta_j^{\text{true}},\psi_j^{\text{true}},k)$
        \ENDIF
    \ENDFOR
    \STATE Store sample $\bigl(\{(I_{j,i},\alpha_i,\beta_i)\}_{i=1}^{20},\,k,\,(\Delta_j^{\text{true}},\psi_j^{\text{true}})\bigr)$ in $\mathcal{D}$
\ENDFOR
\STATE Randomly split $\mathcal{D}$ into $\mathcal{D}_{\text{train}}$ and $\mathcal{D}_{\text{val}}$ with ratio $(1-\nu):\nu$
\STATE\textbf{Model Configuration}
\STATE \textit{Input}: $60$‑channel tensor obtained by stacking $20$ frames $\{(I_i,\alpha_i,\beta_i)\}_{i=1}^{20}$
\STATE \textit{Network}: $f_\theta$ -- FiLM‑conditioned U‑Net
\STATE \textit{Output}: $(\Delta^{\text{pred}},\psi^{\text{pred}})\!\in\!\mathbb{R}^{H\times W}$
\STATE\textbf{Loss (per sample)}
\[
\mathcal{L}_j =\|\Delta_j^{\text{pred}}-\Delta_j^{\text{true}}\|_F^2+
                \|\psi_j^{\text{pred}}-\psi_j^{\text{true}}\|_F^2
\]
\STATE Normalize both channels by $(\pi,\pi/2)$ before computing $\mathcal{L}_j$ 
\STATE\textbf{Training}
\FOR{epoch $=1$ to $K$}
    \FOR{each mini‑batch of size $B$ in $\mathcal{D}_{\text{train}}$}
        \STATE $\mathcal{L}_{\text{batch}}=\frac1B\sum_{j=1}^{B}\mathcal{L}_j$
        \STATE Adam update $\theta\leftarrow\theta-\eta\nabla_\theta\mathcal{L}_{\text{batch}}$
    \ENDFOR
    \STATE Compute validation loss on $\mathcal{D}_{\text{val}}$; apply ReduceLROnPlateau
    \STATE Trigger early‑stopping when no improvement for $P$ epochs
\ENDFOR
\STATE\textbf{Model Selection} $\rightarrow$ $\theta^*$ is the checkpoint with lowest validation loss
\RETURN $\theta^*$
\end{algorithmic}
\end{algorithm}

The training methodology outlined in Algorithm~\ref{alg:supervised-inversion} incorporates several key principles to ensure robust and efficient learning. First, the data synthesis strategy includes a balanced mixture of ideal (noise-free) and physically realistic noisy samples, enabling the network to learn features that are resilient to measurement noise and interlayer crosstalk. This enhances its generalization capability in practical applications. Second, the loss function is carefully designed to balance the contributions of retardance ($\Delta$) and orientation ($\psi$). Since these parameters span different physical ranges, their error terms are normalized by the square of their respective ranges—$1/\pi^2$ for $\Delta$ and $4/\pi^2$ for $\psi$—to ensure equal weighting during optimization and to prevent domination by the parameter with the larger numerical scale. Third, mini-batch gradient updates using the Adam optimizer yield stable and computationally efficient gradient estimates, promoting reliable convergence. To further improve training stability and prevent overfitting, we employ an early stopping criterion based on the validation loss, which halts training when no improvement is observed over a predefined number of epochs. In addition, a ReduceLROnPlateau scheduler is applied to dynamically adjust the learning rate $\eta$ by reducing it when the validation performance plateaus, allowing finer optimization steps and facilitating convergence to a better local minimum.

This data-driven framework supports accurate and noise-resilient birefringence reconstruction across diverse optical modulation states and system configurations. Through FiLM-based conditional modulation, the model dynamically adapts to different physical layers specified by the index $k$, making it suitable for scalable and real-time birefringence analysis in advanced optical storage systems.

\section{Numerical examples}\label{Sec:Num}

We first implemented both the four-frame and five-frame analytical algorithms under varying levels of simulated noise. The results are summarized in TABLE~\ref{tab:errors_with_noise_4_5_frame}, where the reconstruction errors for both the phase retardance $\Delta$ and the slow-axis orientation $\psi$ are evaluated using two norms: the Frobenius norm and the absolute $L^\infty$ norm.

As the noise level increases from $5\%$ to $30\%$, the performance of both frame-based algorithms deteriorates significantly. Even at moderate noise levels (e.g., $15\%$  or $20\%$ ), the relative errors in $\Delta$ and $\psi$ exceed $13$ and $6$, respectively, indicating that both algorithms are highly sensitive to realistic sources of perturbation. Notably, the $L^\infty$ norm errors, which represents worst-case deviations, remain consistently high across all conditions. 

These findings clearly reveal the limitations of traditional analytical approaches, particularly in the presence of noise introduced by interlayer crosstalk, optical imperfections, and detector instability. 

\begin{table}[htbp]
    \centering
    \setlength{\tabcolsep}{8pt} 
    \fontsize{11}{14}\selectfont
    \caption{Errors for $\Delta$ and $\psi$ of Four-frame and Five-frame algorithms under different noise levels.}
    \begin{tabular}{c|cc|cc|cc|cc}
        \hline
        \multirow{2}{*}{Noise} 
        & \multicolumn{4}{c|}{Four-frame} 
        & \multicolumn{4}{c}{Five-frame} \\
        \cline{2-9}
        & \multicolumn{2}{c|}{$\|\cdot\|_F$} 
        & \multicolumn{2}{c|}{$|\cdot|_\infty$}
        & \multicolumn{2}{c|}{$\|\cdot\|_F$} 
        & \multicolumn{2}{c}{$|\cdot|_\infty$} \\
        \cline{2-9}
        & $\Delta$ & $\psi$ & $\Delta$ & $\psi$
        & $\Delta$ & $\psi$ & $\Delta$ & $\psi$ \\
        \hline
        5\%  &    13.4   &  7.3     &   2.4    &  1.2     &    13.0   &   5.7    &   2.4    &    1.3   \\
        10\% &    15.1   &    7.3   &  2.8     &   1.3    &     14.7  &   7.0    &  2.8     & 1.5      \\
        15\% &    16.3   &  7.0     &   2.9    &   1.3    &    16.2   &   6.9    &   2.9    &   1.3    \\
        20\% &    13.4   &   7.4    &   3.1    &   1.5    &    13.1   &    7.2   &  3.1     &    1.5   \\
        25\% &   14.7   &   7.1    &  2.7     &  1.4     &  15.1     &   6.8    &   2.8    &   1.5   \\
        30\% &    16.4    &     7.0  &  2.9     &   1.4    &    16.1   &   7.1    &   2.9    &   1.5     \\
        \hline
    \end{tabular}
    \label{tab:errors_with_noise_4_5_frame}
\end{table}

\subsection{Dataset and training strategy}
The 20-frame FiLM-Conditioned U-Net model is trained and validated on a dataset of $M = 2000$ independent ground-truth birefringence maps $\{(\Delta_j^{\text{true}}, \psi_j^{\text{true}})\}_{j=1}^M$ ($H=W=10$). The values of $\Delta$ and $\psi$ are independently drawn from uniform distributions $\mathcal{U}(0, \pi)$ and $\mathcal{U}(0, \pi/2)$, respectively, to capture the full dynamic range of physical birefringence properties. 
The dataset is randomly partitioned with 85\% (1700 samples) for training and 15\% (300 samples) for validation. For each ground-truth map, one input sample is generated by simulating 20 intensity distributions $I_i$ using a fixed set of 20 distinct $(\alpha_i,\beta_i)$ polarization angle pairs. Among these, 10 are used to generate ideal (noise-free) intensity distributions, and the remaining 10 produce corrupted intensity distributions via a physically motivated noise model. This hybrid construction exposes the network to both clean and noisy conditions, thereby enhancing its robustness to signal degradation.
All intensity distributions are computed via the Jones matrix forward model (Eq.\eqref{eq:cumpImax}) described in Section~\ref{sec:models}. The corrupted images incorporate five degradation sources: phase noise ($\delta_\Delta$), azimuth noise ($\delta_\psi$), multiplicative source fluctuation ($\varepsilon$), and additive Gaussian background noise ($I_{\text{add}}$), each modeled as a zero-mean normal distribution with standard deviation $\sigma=0.05$, as detailed in Table~\ref{tab:parameters}. The total number of layers in the physical model is $N = 20$, with the maximum layer index defined as $L = 20$.
All training and evaluation were conducted using the PyTorch framework on an NVIDIA L40S GPU. The source code is publicly available at \url{https://github.com/z1998w/5D-optical-storage}.


\begin{table}[htbp]
\centering
\caption{Parameters required for simulating intensity $I$ in the forward model Eq.~\eqref{eq:computeIwithoutnoise}}
\begin{tabular}{|c|c|c|c|}
\hline
\textbf{Parameter Name} & \textbf{Symbol} & \textbf{Value/Distribution} & \textbf{Description} \\
\hline
\hline
Retardance magnitude & $\Delta$ & $\sim \mathcal{U}(0, \pi)$ & \makecell{optical phase delay of the \\ voxel} \\
\hline
Slow-axis orientation & $\psi$ & $\sim \mathcal{U}(0, \pi/2)$ & optical slow-axis azimuth \\
\hline
\makecell[c]{Parameter noise \\ (retardance)} & $\delta_\Delta$ & \makecell[c]{$\sim \mathcal{N}(0, \sigma_{\Delta}^2)$ \\ with\,\, $\sigma_{\Delta}=0.05$ } & noise level in phase retardance \\
\hline
\makecell[c]{Parameter noise \\ (azimuth)} & $\delta_\psi$ & \makecell[c]{$\sim \mathcal{N}(0, \sigma_{\psi}^2)$ \\ with\,\, $\sigma_{\psi}=0.05$} & noise in slow-axis direction \\
\hline
Multiplicative noise & $\varepsilon$ & \makecell[c]{$\sim \mathcal{N}(0, \sigma_{\text{scr}}^2)$ \\ with \, $\sigma_{\text{scr}}=0.05$ } & \makecell{source fluctuation} \\
\hline
Additive noise & $I_{\text{add}}$ & \makecell[c]{$\sim \mathcal{N}(0, \sigma_{\text{add}}^2)$ \\ with\, $\sigma_{\text{add}}=0.05$} & \makecell{detector dark current} \\
\hline
Polarizer angle & $\alpha$ & e.g., $90^\circ$ & \makecell{analyzer polarizer rotation \\ angle} \\
\hline
Retarder angle & $\beta$ & e.g., $180^\circ$ & \makecell{phase modulation retarder \\ angle} \\
\hline
Light transmittance factor & $\tau$ & 0.98 & light transmission coefficient \\
\hline
Background intensity & $I_{\text{min}}$ & 0.001$I_{0}$ & background signal level \\
\hline
\end{tabular}
\label{tab:parameters}
\end{table}
The synthetic dataset and algorithmic framework are constructed upon a physically faithful multilayer Jones matrix model, simulating realistic light-matter interactions in 5D optical data storage. As shown in TABLE~\ref{tab:Imax_parameters}, the incident light field propagates through $N$ birefringent layers with interlayer spacing $h=0.2$, semi-aperture angle $\theta=20^\circ$, and grid spacing $\omega=0.1$. Each voxel may encode an individual $2\times 2$ Jones matrix $\boldsymbol{J}_g$, drawn from a space of random unitary or birefringent matrices, representing the encoded optical modulation from nanogratings. The grating density ratio $d$ controls the fractional occupancy of birefringent elements within each layer (e.g., $d=0.02$). The physical propagation, modulation, and projection to intensity values are modeled end-to-end, thereby enabling the neural network to learn directly from physically grounded measurements.

\begin{table}[htbp]
\centering
\caption{Parameters required for computing $I_{\text{max}}$ from Jones matrix model Eq.~\eqref{eq:cumpImax}}\label{tab:Imax_parameters}
\begin{tabular}{|c|c|c|c|}
\hline
\textbf{Parameter Name} & \textbf{Symbol} & \textbf{Value/Distribution} & \textbf{Description} \\
\hline
\hline
Number of layers & $N$ & e.g., $100\sim 200$ & \makecell{total number of storage layers} \\
\hline
Grating density ratio & $d$ &  0.02 & \makecell{fraction of nanogratings} \\
\hline
Interlayer spacing & $h$ & 0.2 & interlayer spacing \\
\hline
Semi-aperture angle & $\theta$ & $20^\circ$ & semi-aperture angle \\
\hline
Grid spacing & $\omega$ & 0.1 & grid spacing \\
\hline
Modulation weight factor & $\sigma(n)$ & $\displaystyle{\sigma(n) \propto \frac{1}{|n-k|+1}}$ & \makecell{Distance-dependent \\ modulation strength} \\
\hline
\makecell{Random Jones matrix \\ for grating $g$} & $\boldsymbol{J}_g$ & \makecell{random $2\times2$ unitary \\ or birefringent matrices} & \makecell{modulation from \\ individual nanogratings} \\
\hline
\end{tabular}
\end{table}



Optimization is performed using the Adam optimizer with a batch size of $B=8$ for a maximum of $K=1000$ epochs, starting with an initial learning rate of $\eta=3 \times 10^{-4}$. The training process is governed by two key control mechanisms based on the validation loss, which is monitored at the end of each epoch. First, a `ReduceLROnPlateau` learning rate scheduler is applied, which halves the learning rate if the validation loss shows no improvement for 20 consecutive epochs. Second, an early stopping mechanism with a patience of $P=500$ epochs is employed to prevent overfitting; training is terminated if the validation loss does not decrease for this duration. The final model selected for evaluation is the one corresponding to the epoch with the lowest achieved validation loss.

For performance evaluation, we construct a dedicated test dataset based on 20 previously unseen ground-truth $(\Delta, \psi)$ maps. Each of these maps is used to generate two complementary test sets. The first is an ideal test set comprising noise-free intensity distributions $I$. The second is a noisy test set designed for robustness evaluation, where each ground-truth map is used to generate a set of corrupted intensity distributions $I$ at all 20 layers. This dual-test setup enables both unbiased generalization assessment and detailed layer-wise analysis of noise robustness.

\subsection{Results and Analysis}
\label{Results}

\begin{figure}[htbp] 
\centering
\subfigure{
\includegraphics[width=0.4\linewidth]{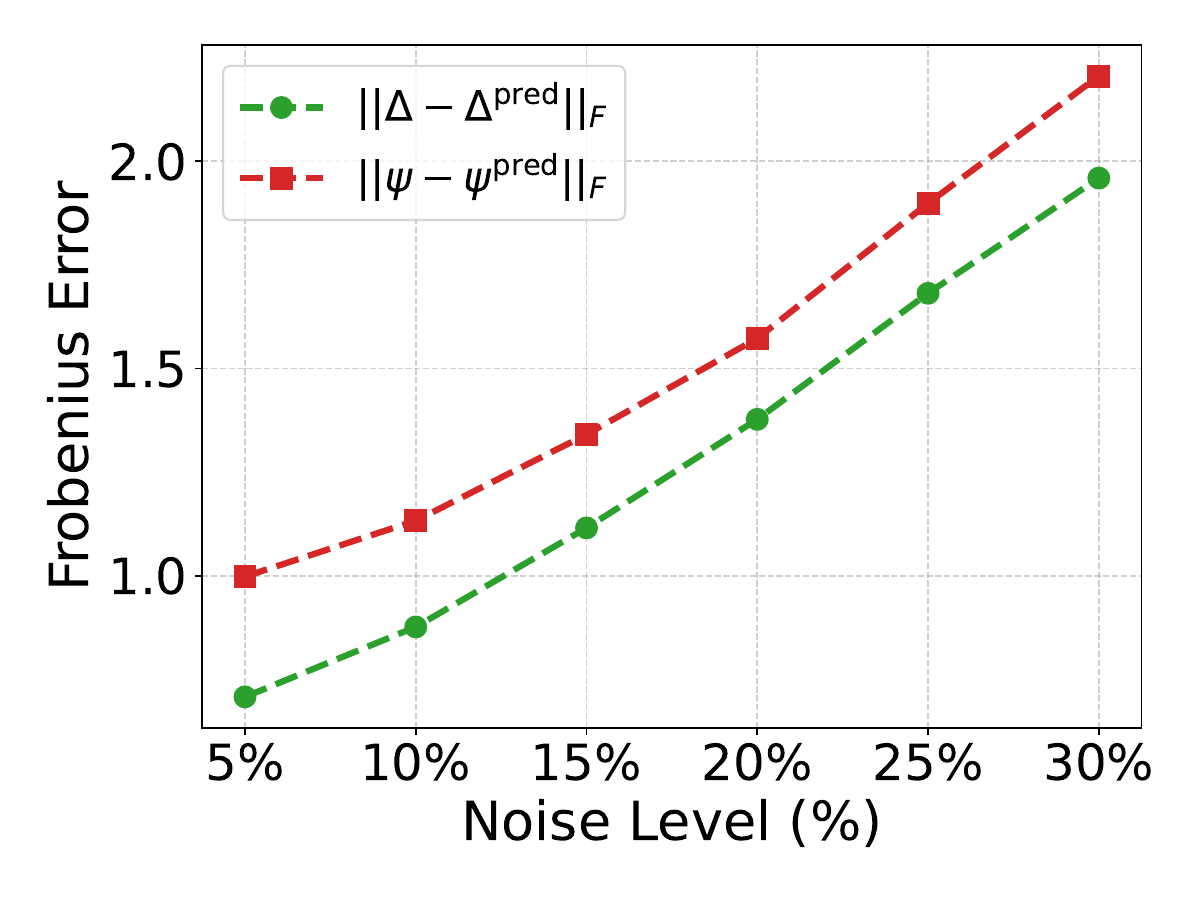}
\label{fig:Noise_F_error}
}
\hspace{0.05\linewidth}
\subfigure{
\includegraphics[width=0.4\linewidth]{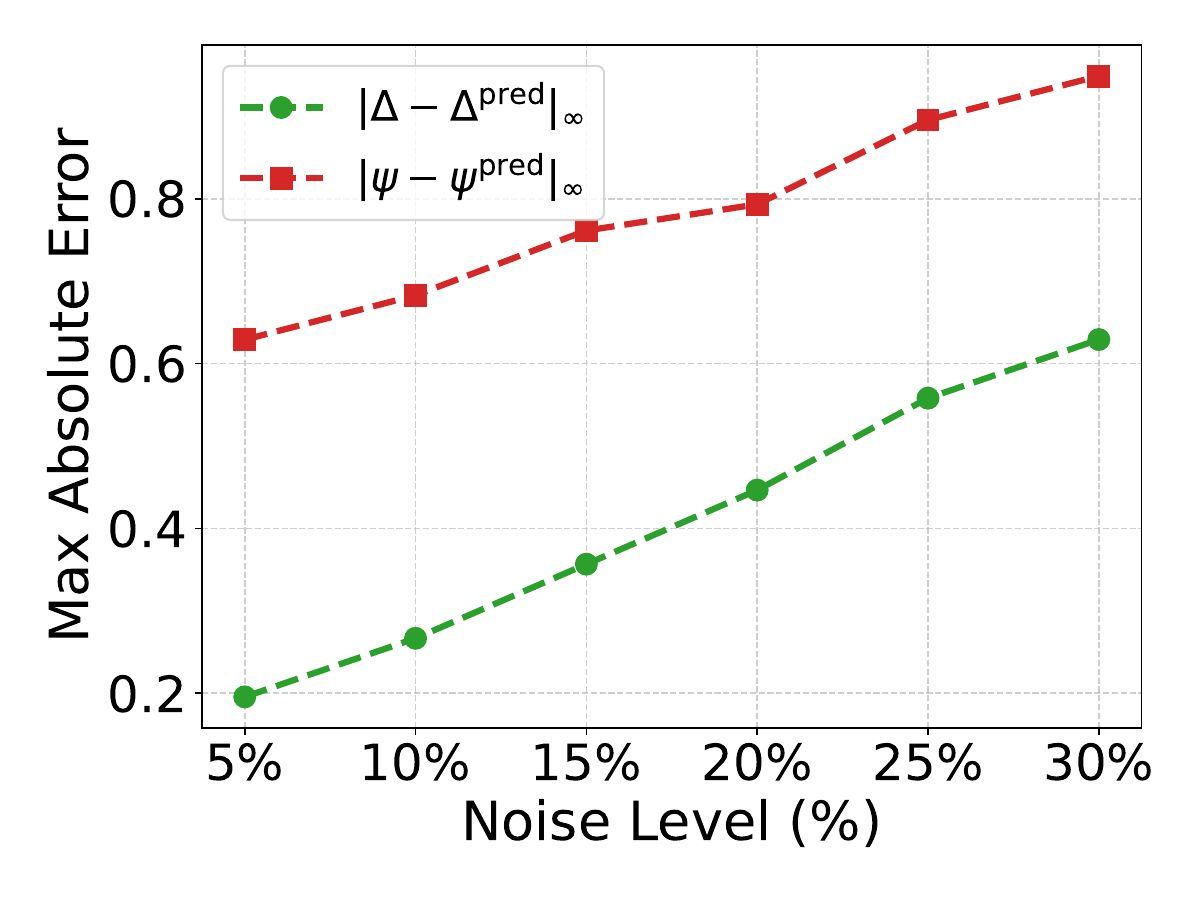}
\label{fig:Noise_max_error}
}
\caption{Mean errors for $\Delta$ and $\psi$ over 20 samples and all layer indices $k$ under varying noise levels in the test set.
}
\label{fig:Layer_noise_VS_error}
\end{figure}

The proposed 20-frame FiLM-Conditioned U-Net model's reconstruction accuracy and robustness are evaluated using two primary metrics: the Frobenius norm ($\| \cdot \|_F$) for overall error and the infinity norm ($|\cdot|_{\infty}$) for worst-case pixel deviation. The results, summarized in TABLE~\ref{tab:mean_errors_over_20_samples_all_k}, \ref{tab:errors_with_and_without_noise}, and \ref{tab:error_different_layers_single_sample}, and visualized in FIG.~\ref{fig:Layer_noise_VS_error}, collectively demonstrate the model's superior performance.

The performance of the model on the noise-free test set is highly accurate, as shown in the averaged results in TABLE~\ref{tab:mean_errors_over_20_samples_all_k}. It achieves mean Frobenius errors of 0.466 for $\Delta$ and 0.84 for $\psi$, confirming its capability to correctly learn the complex, non-linear inverse mapping from intensity distributions to birefringence parameters.

More critically, the model demonstrates excellent robustness when subjected to realistic physical noise. When comparing the noise-free and noisy test sets in TABLE~\ref{tab:mean_errors_over_20_samples_all_k}, the average Frobenius error for $\Delta$ shows only a moderate increase from 0.466 to 0.707, while the error for $\psi$ increases from 0.814 to 1.01. Similarly, the mean worst-case pixel errors ($|\cdot|_\infty$) remain well-contained. Such robustness is further illustrated in TABLE~\ref{tab:errors_with_and_without_noise}, where for 5 samples with noise originating from layer $k=12$, the error metrics increase only marginally. This indicates that the learned inverse map is not easily disturbed by the combined effects of signal corruption and measurement uncertainties. TABLE~\ref{tab:error_different_layers_single_sample} confirms that the model is insensitive to the defect's physical location, as the error for a fixed sample remains stable even when the noise-inducing layer $k$ is varied throughout the stack, highlighting the effectiveness of the FiLM conditioning mechanism.

FIG.~\ref{fig:Layer_noise_VS_error} evaluates the robustness of the model under noise levels progressively increasing from 0.05 to 0.30.  The Frobenius errors for $\Delta$ rise from below $0.8$ to roughly $2.0$, while those for $\psi$ increase from under $1.0$ to about $2.5$ at the highest noise level. Although the errors grow with noise, the growth rate remains moderate and well-controlled, without signs of abrupt escalation. The corresponding maximum absolute errors follow a similar pattern, staying below $0.8$ for $\Delta$ and under $1$ for $\psi$ even at the highest corruption level. This controlled increase indicates that the model maintains predictable and stable performance, avoiding catastrophic degradation even under substantial physical noise.

\begin{table}[htbp]
    \centering
    \setlength{\tabcolsep}{11pt} 
    \fontsize{11}{13}\selectfont
    \caption{Mean errors for $\Delta$ and $\psi$ averaged over 20 samples and all layer indices $k$ using the 20-frame FiLM-Conditioned U-Net (Algorithm~\ref{alg:supervised-inversion}).}
    \begin{tabular}{c |c c c c}
        \hline
         & $\|\Delta-\Delta^{\text{pred}}\|_F$ & $\|\psi-\psi^{\text{pred}}\|_F$ & $|\Delta-\Delta^{\text{pred}}|_\infty$ & $|\psi-\psi^{\text{pred}}|_\infty$ \\
        \hline
        Noise-free Test Set & 4.66e-01 & 8.14e-01 & 1.31e-01 & 5.58e-01 \\
        Noisy Test Set      & 7.07e-01 & 1.01e+00 & 1.98e-01 & 6.46e-01 \\
        \hline
    \end{tabular}
    \label{tab:mean_errors_over_20_samples_all_k}
\end{table}

\begin{table}[htbp]
    \centering
    \setlength{\tabcolsep}{12pt} 
    \fontsize{11}{13}\selectfont
    \caption{Errors for $\Delta$ and $\psi$ on 5 samples under noise-free and noisy test sets.}
    \begin{tabular}{ccccc}
        \hline
        Sample & $\|\Delta-\Delta^{\text{pred}}\|_F$&$\|\psi-\psi^{\text{pred}}\|_F$ & $|\Delta-\Delta^{\text{pred}}|_\infty$ & $|\psi-\psi^{\text{pred}}|_\infty$\\
         \hline
        \multicolumn{5}{c}{Noise-free Test Set}\\
        \hline
        1 & 5.01e-01 & 7.90e-01 & 1.85e-01 & 4.72e-01 \\
        2 & 4.70e-01 & 7.30e-01 & 1.46e-01 & 4.77e-01 \\
        3 & 4.99e-01 & 5.44e-01 & 1.31e-01 & 4.06e-01 \\
        4 & 4.75e-01 & 2.96e-01 & 1.48e-01 & 1.25e-01 \\
        5 & 4.64e-01 & 9.74e-01 & 1.03e-01 & 5.94e-01 \\
        \hline
        \multicolumn{5}{c}{Noisy Test Set (Layer $k = 12$)}\\
        \hline
        1 & 1.16e+00 & 1.05e+00 & 3.42e-01 & 6.71e-01 \\
        2 & 6.37e-01 & 9.74e-01 & 2.24e-01 & 2.71e-01 \\
        3 & 6.75e-01 & 7.81e-01 & 1.86e-01 & 4.93e-01 \\
        4 & 9.47e-01 & 7.09e-01 & 2.81e-01 & 2.44e-01 \\
        5 & 6.88e-01 & 1.12e+00 & 1.83e-01 & 8.29e-01 \\
        \hline
    \end{tabular}
    \label{tab:errors_with_and_without_noise}
\end{table}

\begin{table}[htbp]
    \centering
    \setlength{\tabcolsep}{12pt} 
    \fontsize{11}{13}\selectfont
    \caption{Errors for $\Delta$ and $\psi$ under noisy test sets on a fixed sample across different layer indices $k$.}
    \begin{tabular}{c c c c c}
        \hline
        Layer $k$ & $\|\Delta-\Delta^{\text{pred}}\|_F$ & $\|\psi-\psi^{\text{pred}}\|_F$ & $|\Delta-\Delta^{\text{pred}}|_\infty$ & $|\psi-\psi^{\text{pred}}|_\infty$ \\
        \hline
        1 & 1.10e+00 & 8.29e-01 & 2.80e-01 & 6.06e-01 \\
        5  & 5.94e-01 & 9.77e-01 & 2.44e-01 & 5.42e-01 \\
        11 & 5.73e-01 & 9.97e-01 & 1.60e-01 & 7.04e-01 \\
        17 & 7.12e-01 & 9.99e-01 & 1.91e-01 & 8.08e-01 \\
        20 & 7.19e-01 & 7.61e-01 & 2.23e-01 & 4.66e-01 \\
        \hline
    \end{tabular}
    \label{tab:error_different_layers_single_sample}
\end{table}

\section{Conclusion}\label{Sec:Con}

We validated the proposed modeling and inversion approach through both simulations and preliminary experimental studies on femtosecond laser-written nanograting voxel arrays. The results demonstrate that the multilayer Jones matrix model accurately captures the polarization-dependent signal variations in multi-layer 5D optical storage systems. Furthermore, the proposed 20‑frame FiLM‑Conditioned U‑Net reconstruction algorithm achieved over an order-of-magnitude reduction in retrieval error compared with conventional frame-based methods, even under strong interlayer crosstalk and noise. 
In particular, quantitative evaluations show that the mean Frobenius and absolute errors for phase retardance were reduced from approximately 15 to below 1.5, and from about 3 to below 0.3, respectively. For slow-axis orientation, the mean Frobenius and absolute errors decreased from roughly 7 to around 0.7, and from about 1.5 to around 0.5, respectively.
These findings suggest that combining physical modeling with data-driven inversion strategies is a promising direction for enhancing the readout fidelity of 5D optical storage systems.

While this study provides an initial demonstration of the approach, further validation on larger-scale experimental datasets and more complex voxel encoding schemes is required to fully establish generalization and practical utility of the method. The proposed architecture is inherently scalable: although the present experiments use 20-frame inputs, it can be extended to a larger number of frames and adapted to diverse data acquisition strategies, with higher-capacity networks expected to yield further gains in reconstruction accuracy. Overall, integrating forward modeling with deep learning offers a promising and flexible framework for enhancing the robustness and scalability of polarization-based optical data storage technologies.

\begin{acknowledgments}
This work was partially supported by the National Key Research and Development Program of China (No. 2022YFC3310300), Shenzhen Sci-Tech Fund Grant No. RCJC20231211090030059), National Natural Science Foundation of China (No. 12171036) and Beijing Natural Science Foundation (No. Z210001). 
\end{acknowledgments}


\bibliographystyle{plain}
\bibliography{ref}

\end{document}